\documentclass[preprint,12pt]{elsarticle}

\usepackage{amssymb}
\usepackage{srcltx}
\usepackage{amsmath}

\usepackage{graphicx}% Include figure files
\usepackage{dcolumn}% Align table columns on decimal point
\usepackage{bm}% bold math
\usepackage{multicol}

\usepackage{amsthm}

\newtheorem*{theorem}{Theorem}

\journal{}

\begin{document}

\begin{frontmatter}

\title{Construction of $N$th-order rogue wave solutions for Hirota equation by means of bilinear method}
\author[rvt]{Gui Mu}\author[els]{Zhenyun Qin\corref{cor1}}\ead{zyqin@fudan.edu.cn}
\cortext[cor1]{Corresponding author}
 \address[rvt]{College of Mathematics and Information Science, Qujing Normal University, \\  Qujing 655011, PR China}
 \address[els]{School of Mathematics, Key Laboratory of Mathematics for Nonlinear Science \\ and Shanghai Center for Mathematical Sciences, Fudan University, \\ Shanghai 200433, PR China}

%\title{}
%\author{Gui Mu $^{1}$,\ Zhenyun Qin $^{2}$\footnote{Corresponding author E-mail address: zyqin@fudan.edu.cn} \\
%\it\small{{$ ^{1}$ College of Mathematics and Information Science, Qujing Normal University,}}\\
%\it\small{{ Qujing 655011, PR China}}\\
%\it\small{{ $^{2}$
%School of Mathematics and Key Laboratory of Mathematics for Nonlinear Science,}}\\
% \it\small{{ Fudan University, Shanghai 200433, PR China}}}
%\date{}
%\maketitle
%\vspace{0.5cm}

\begin{abstract} In this work, we focus on the construction of $N$th-rouge wave solutions for the Hirota equation by utilizing the bilinear method. The formula can be represented in terms of determinants. In addition, some interesting dynamic patterns of rogue waves are exhibited.
\end{abstract}
%They can be reduced to those for nonlinear Schr\"{o}dinger equation corresponding to the reference \cite{oh}.

\end{frontmatter}

%\textbf{AMS subject classification.} 22E46, 53C35, 57S20
%\begin{center}\setcounter{equation}{0} \vspace{1ex} \indent\end{center}

\section{Introduction}

In recent years, rogue waves have been intensively aroused much attention in many fields from optics to Bose-Einstein condensates \cite{kh}-\cite{pe}. Mathematically, the rational solution (so called Peregrine solution) of focusing nonlinear Schr\"{o}dinger equation (NLSE) is initially used to explain such rogue wave phenomenon in these contexts. This mainly stems from it localized in both coordinates and reaches a height which is three times of background amplitude. Starting with the pioneer work made by Akhmediev research group \cite{aa1}, various technique have been taken to derive the \emph{N}th-order rational solutions of NLSE \cite{g1}-\cite{mu2014}. Indeed, the construction of rational solutions requires some special methods that differ from those utilized to get multisoliton solutions \cite{aa4}. Currently, the Darboux transformation method and the bilinear method are regarded as two effective methods since they are used to create the whole hierarchy of rational solutions for a wide class of nonlinear evolution equations \cite{aa2}-\cite{mu3}.

Hirota equation is a modified NLSE especially when the third order dispersion and time-delay changes are taken into account. With regard to rogue wave solutions of Hirota equation, they have been announced using Darboux transformation method \cite{aa2}-\cite{he}. Recalling that the results obtained by bilinear method is more explicit than Darboux transformation in the framework of NLSE \cite{oh}, a question arises whether the rogue wave solutions of Hirota equation can be achieved with the help of bilinear method like NLSE. Here, we give a positive answer. % This is the purpose of this work.

\section{General rogue wave solutions of Hirota equation}

We consider the Hirota equation in the following form
\begin{eqnarray}\label{1}
&&i\phi_t+\phi_{xx}+2|\phi|^2\phi+iv\phi_{xxx}+6iv|\phi|^2\phi_x=0,
\end{eqnarray}
Since rogue waves are always assumed to approach a same constant background when $t\rightarrow \pm \infty$, let us consider the variable transformation
\begin{eqnarray}\label{2}
&&\phi=e^{2it}u,
\end{eqnarray}
then the Hirota equation (\ref{1}) becomes
\begin{eqnarray}\label{3}
iu_t+u_{xx}+2(|u|^2-1)u+ivu_{xxx}+6iv|u|^2u_x=0,
\end{eqnarray}
Using the independent variable transformation
\begin{eqnarray}\label{6}
u=\frac{g}{f},
\end{eqnarray}
where $f$ is a real function and $g$ is a complex one with respect to variables $x$ and $t$.  Then Eq.(\ref{3}) is transformed into the following bilinear form
\begin{eqnarray}
&& \label{5} (D_x^2+2)f\cdot f=2g\overline{g},\\ \label{6}
&& (iD_t+D_x^2+ivD_x^3+6ivD_x)g\cdot f=0,
\end{eqnarray}
where the $D$ is Hirota's bilinear differential operator (see \cite{oh}).  Hereafter,  `-' represents complex conjugation.

Just like NLSE \cite{oh}, by applying bilinear equations (\ref{5}) and (\ref{6}), $N$th-order rational solution of Hirota equation (\ref{1}) can be constructed by the following theorems.
\begin{theorem} The Hirota equation (\ref{1}) has rational solutions
\begin{eqnarray}\label{7}
\phi_N(x,t)=e^{2it}\frac{\tau_{1}}{\tau_{0}},
\end{eqnarray}
where $\tau_{n}=\det\limits_{1\leq i,j \leq N}\left(m_{2i-1,2j-1}^{(n)} \right)$ and the matrix elements are given by
\begin{eqnarray}
&& m_{i,j}^{(n)}=\nonumber \sum^{i}_{k=0}\frac{a_k}{(i-k)!}(p \partial p+\xi+n)^{i-k}\sum^{j}_{l=0}\frac{\overline{a}_l}{(j-l)!}(q\partial q +\eta-n)^{j-l}\frac{1}{p+q}\bigg|_{p=1, q=1},
\end{eqnarray}
where
\begin{eqnarray*}
&&\xi=px+(2i{p}^{2}-3v{p}^{3}-3vp)t,\\
&&\eta=qx-(2i{q}^{2}+3v{q}^{3}+3vq)t,
\end{eqnarray*}
Here, $a_k (k=1\cdot\cdot\cdot N)$ are the arbitrary complex constants and $i, j$ are  positive integers.
\end{theorem}

Similar to the analysis of NLSE \cite{oh}, we may also set $a_0=b_0=1, a_2=a_4=a_6=\cdots=b_2=b_4=\cdots=0$ without loss of generality and remain the irreducible complex parameters $a_3, a_5, \cdots, a_{2N-1}$. Furthermore, these rational solutions could also be expressed in a more explicit form in terms of Schur polynomial.

\section{Dynamics of rogue waves}

In this section, we give some examples to illustrate the dynamics of rouge waves in the Hirota equation by applying the above Theorem.

Firstly, setting $N=1$ in Theorem will produce the following first order rogue wave solution
\begin{eqnarray}
\phi_1(x,t)=e^{2it}\frac{m_{11}^{(1)}}{m_{11}^{(0)}},
\end{eqnarray}
where
\begin{eqnarray*}
&&m_{11}^{(0)}=\frac{1}{2}(x+(2i-6v)t+a_1-\frac{1}{2})(x-(6v+2i)t+\overline{a}_1-\frac{1}{2})+\frac{1}{8},\\
&&m_{11}^{(1)}=\frac{1}{2}(x+(2i-6v)t+a_1+\frac{1}{2})(x-(6v+2i)t+\overline{a}_1-\frac{3}{2})+\frac{1}{8},
\end{eqnarray*}
The above algebra structure is extremely similar to the formula in NLSE.  Let $f=m_{11}^{(0)}$ and $g=m_{11}^{(1)}$, direct substitution shows that $f$ and $g$ are solutions of bilinear equations (\ref{5}) and (\ref{6}). After setting $a_1=0$,
then we could obtain the first order rogue wave solution
\begin{eqnarray}\label{9}
\phi_1=e^{2it}(1-\frac{4(1+4it)}{1+4(\widetilde{x}-6vt)^2+16t^2}),
\end{eqnarray}
where $\widetilde{x}=x-\frac{1}{2}$. In (\ref{9}), take $v=0$ and let $t\rightarrow-t$, then this solution is reduced to the first order rogue wave solution of NLSE \cite{oh}.

In order to get second order rogue wave solution of equation (\ref{3}), we take $N=2$ in Theorem. According to the previous analysis, we may set $a_1=a_2=0$ in (\ref{7}), this case leads to
%\begin{eqnarray}
%u(x,t)=\frac{\left|
%                           \begin{array}{cc}
%                             m_{11}^{(1)} & m_{13}^{(1)} \\
%                             m_{31}^{(1)} & m_{33}^{(1)}
%                           \end{array}
%                         \right|
%}{\left|
%                           \begin{array}{cc}
%                             m_{11}^{(0)} & m_{13}^{(0)} \\
%                             m_{31}^{(0)} & m_{33}^{(0)}
%                           \end{array}
%                         \right|},
%\end{eqnarray}
%So the explicit expressions of second order rogue wave solution of Hirota equation (\ref{1}) is derived as
 \begin{eqnarray*}
\phi_2=e^{2it}(1+\frac{G}{F}),
 \end{eqnarray*}
where
\begin{eqnarray*}
&&F=16\,{x}^{6}-48\,{x}^{5}+ 12(24t^2-3-64t^4)x+24 (3+8t^2){x}^{4}+1024\,{t}^{6}\\
&&\qquad-24(3+16t^2) {x}^{3}+1920\,{t}^{4}+288\,{t}^{2}+ 24(3+32t^4) {x}^{2}+9+\Lambda_1
\\
&&\qquad +24(a_3+\overline{a}_3)(432\,{t}^{3}{v}^{3}+\alpha_1 {v}^{2}+ \alpha_2v-2\,{x}^{3}+3\,{x}^{2}-12\,{t}^{2}+24\,{t}^{2}x\\
&&\qquad +48i(\overline{a}_3-a_3)t[3+8\,{t}^{2}+72v\,t(x-3vt-\frac{1}{2})+6\,x(1-x)]+144|a_3|^2,\\
&&G=24({x}^{3}(4it+1)(4-2x)+ 3(1-4\,it)x+16t^2(4\,i{t}+3)x(1-x)-6{x}^{2}\\
&&\qquad+12\,it-32\,i{t}^{3}-128\,i{t}^{5}-48
\,{t}^{2}-160\,{t}^{4}+\Lambda_2\\
&&\qquad+6a_3(36\,{t}^{2}{v}^{2}+ 12t(1+2\,i{t}-x) v-4\,{t}
^{2}-4\,ixt+4\,it+1-2\,x+{x}^{2})
\\&&\qquad+6\overline{a}_3[-36\,{t}^{2}{v}^{2}+ 12t(2\,i{t}+x) v-4\,ixt-{x}^{
2}+4\,{t}^{2}]),
\end{eqnarray*}
with
\begin{eqnarray*}
&&\Lambda_1=10368{t}^{4}\left[36\,vt(2{t}{v}-2x+1)+24\,{t}^{2}+30
\,{x}^{2}-30\,x+1 \right]{v}^{4} \\
&&\qquad -1728{t}^{3} \left( 3-48\,{t}^{2}+40\,{x}^{3}+96\,{t}^{2}x+12\,x-60\,{x}^{2} \right) {v}^{3}\\
&&\qquad +288{t}^{2} \left[17-3\,x+144\,{x}{t}^{2}(x-1)+
96\,{t}^{2}({t}^{2}+1)+30\,{x}^{2}(x-1)^2 \right] {v}^{2}\\
&&\qquad -24t [ 192\,{t}^{2}({x}^{3}-t^2+x+2t^2x-\frac{3}{2}x^2-\frac{1}{8})+56\,{x}^{3}-30\,{x}^{2}\\
&&\qquad-60\,{x}^{4}+24\,{x}^{5}+
36\,x-9 ]v,
\end{eqnarray*}
\begin{eqnarray*}
&&\Lambda_2=864(2\,x -3v\,t-1) (1+4i\,t){t}^{3}{v}^{3}-72t^2[8t(4\,{t}^{2}-3\,x+3\,{x}^{2}+2) i\\
&&\qquad+6\,{x}^{2}+24\,{t}^{2}+7-6\,x
]v^2+6[4t(1+4\,i{t}){x}^{2}(2x-3)\\
&&\qquad +4t( 32\,i{t}^{3}+24\,{t}^{2}+8\,i{t}+5
 ) x-64\,i{t}^{4}-48\,{t}^{3}-4\,i{t}^{2}-7\,t
]v,\\
&&\alpha_1=-108(2\,x -1) {t}^{2},\quad \alpha_2=12\,t ( 3\,{x}^{2}-3\,x-12\,{t}^{2}-2).
\end{eqnarray*}
Direct substitution shows that $\phi_2$ is indeed the solution of Hirota equation (\ref{1}) using the symbolic software $Maple$.  If we take $v=0$ and make the transformation $t\rightarrow -t$, the above solution will be reduced to the second order rogue wave solution of NLSE presented by Ohta and Yang \cite{oh}. Moreover, it is found that the maximum of $|\phi_2(x,t)|$ is not affected by the values of $v$ and achieved $5$ when $a_3=-\frac{1}{12}$. This means the value of $v$ only contributes to the skewness of rogue waves. In the following, to show the effectiveness of our method, we mainly study the third order case graphically. Figure 1
(a) and (c) display two kinds of asymmetrical third order rogue waves while Figure 1 (b) gives a symmetrical one. In the symmetrical case, the maximum of $|\phi_3(x,0)|$ reach 7. In the two asymmetrical cases, from the central line $|\phi_3(x,0)|$ (see Figure 1d), we observe that all the zeros of $|\phi_3(x,0)|$  occur on the real $x$ axis but the maximum of $|\phi_3(x,0)|$ is lower than 7. It indicates that the spatial-temporal symmetry may be a criterion to identify the generation of higher order rogue waves. In addition, by some different choices of the free parameter $a_3$, we plot four kinds of typical triangular rogue wave cascades patterns in Figure 2. It is evident that the spatial-temporal distribution of the rouge wave rotates an angle $\pi$ around its central position when we change the sign of the value of $a_3$.

\begin{figure*}
  \centering
   %Requires \usepackage{graphicx}
  \includegraphics[width=5.6in]{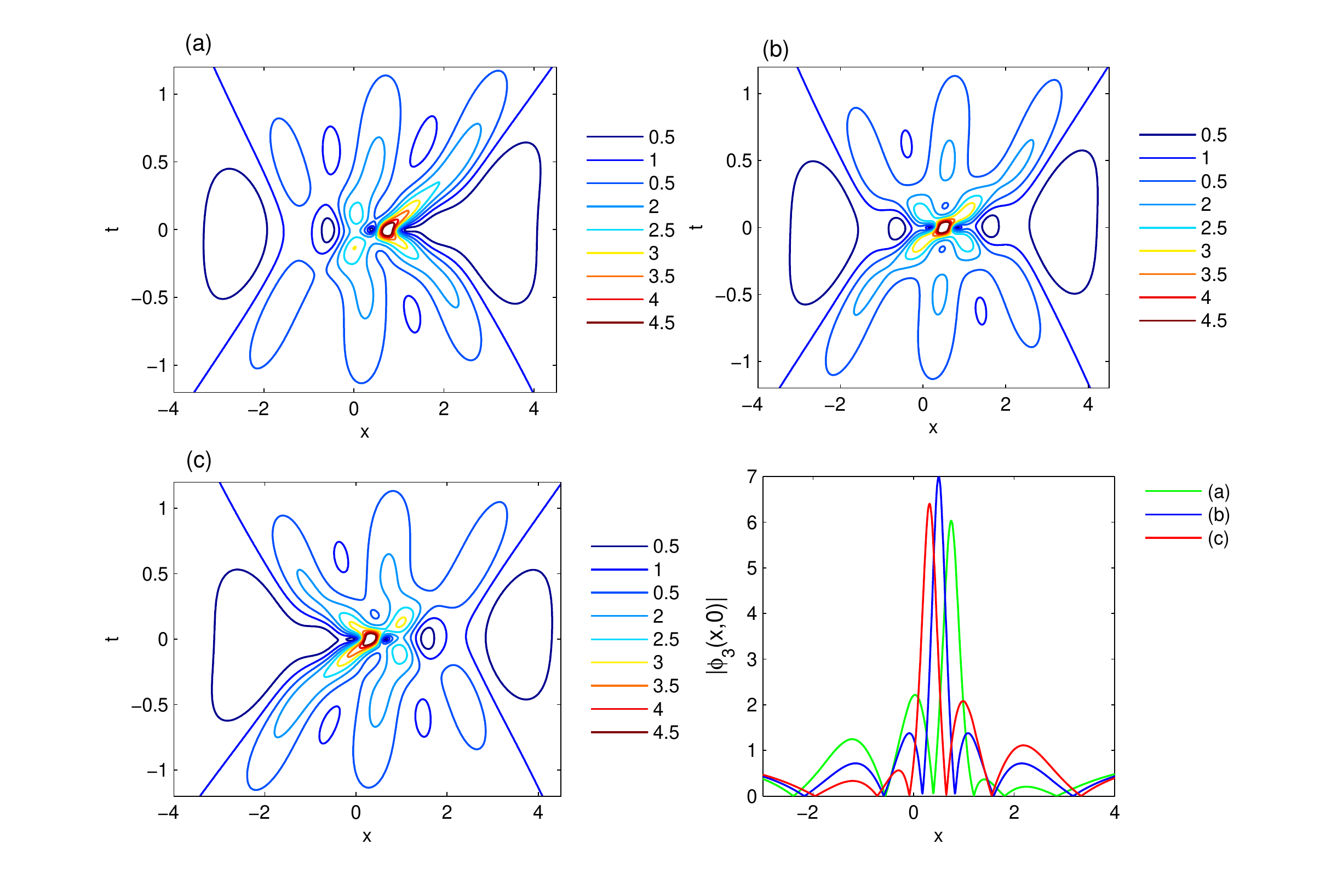}\\
  \caption{ (Color online) The isolines of third order rogue wave solution of Hirota equation for parameters: (a) $v=\frac{1}{20}, a_3=-\frac{1}{5}, a_5=-\frac{1}{240}$, (b) $v=\frac{1}{20}, a_3=-\frac{1}{12}, a_5=-\frac{1}{240}$, (c) $v=\frac{1}{20}, a_3=-\frac{1}{50000}, a_5=-\frac{1}{240}$. Plotted in (d) is the functions $|\phi_3(x,0)|$ corresponding to (a)-(c).}
\end{figure*}

\begin{figure*}
  \centering
   %Requires \usepackage{graphicx}
  \includegraphics[width=6.0in]{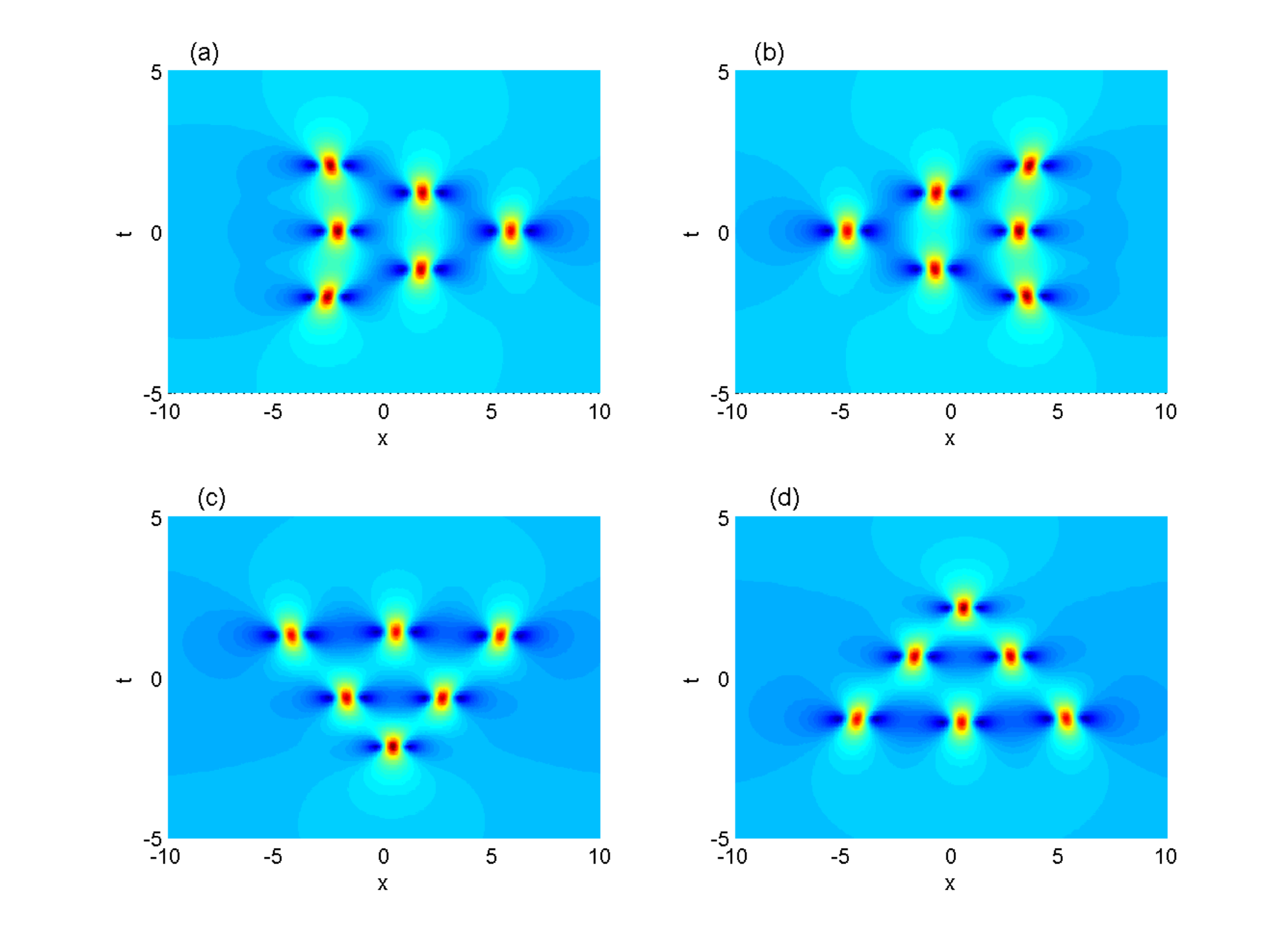}\\
  \caption{ (Color online) Spatial-temporal distribution of third order rogue wave solution in  Hirota equation for parameters $v=\frac{1}{200}, a_5=0$: (a) $a_3=\frac{25}{3}$, (b) $a_3=-\frac{25}{3}$, (c) $a_3=\frac{25}{3}i$. (d) $a_3=-\frac{25}{3}i$.}
\end{figure*}

\section{Conclusions}
To sum up, it has been shown that general $N$th-order rogue wave solution of Hirota equation can be produced with the help of bilinear method. These solutions are expressed in terms of determinants and possess similar structures like NLSE in \cite{oh}. The truth of these results can be testified by symbolic software \emph{Maple} using the bilinear equations (\ref{5}) and (\ref{6}). Taking third order rogue wave solutions as a example and adjusting the values of the free irreducible parameters, we mainly investigate some higher order rogue wave patterns and triangular rogue wave cascades patterns.  These results demonstrate that bilinear method is a powerful method to construct rogue waves of nonlinear evolution equations and we expect the rogue wave solutions of derivative NLSE may also be established by means of bilinear method. This topic is left to future studies.

\section{Acknowledgments}

This work is sponsored by Shanghai Pujiang Program (No. 14PJD007)
and the Natural Science Foundation of Shanghai (No. 14ZR1403500 ), Shanghai Center for Mathematical Sciences and the Young
Teachers Foundation (No. 1411018) of Fudan university and Yunnan province
 project Education Fund (No. 2013C012).

\end{document}